\documentclass{JINST}

\title{Benchmarking TPB-coated Light Guides for Liquid Argon TPC Light Detection Systems}

\author{B. Baptista$^1$, L. Bugel$^2$, C. Chiu$^2$,  J.M. Conrad$^2$,
  C.M. Ignarra$^2$,  B.J.P. Jones$^2$, T. Katori$^2$, S. Mufson$^1$\\
\llap{$^1$}Astronomy Dept., Indiana University, , Bloomington, Indiana, 47405\\
\llap{$^2$}Physics Dept., Massachusetts Institute of Technology, Cambridge, MA 02139\\}

\abstract{
  Scintillation light from liquid argon is produced at 128 nm and thus must be shifted 
to visible wavelengths in light detection systems used 
  for Liquid Argon Time Projection Chambers (LArTPCs).  To date, 
  designs have employed tetraphenyl butadiene (TPB) coatings on photomultiplier 
  tubes (PMTs) or plates placed in front of the PMTs.   Recently, a new approach
  using  TPB-coated light guides was proposed.    In this paper, we 
 show that the response of lightguides coated with TPB in a UV Transmitting (UVT)
  acrylic matrix is very similar to that of a coating using a
  polystyrene (PS) matrix.  We obtain a factor of three higher light yield 
  than has been previously reported
  from lightguides. This paper provides 
  information on the response of the lightguides so that these can be
  modeled in simulations for future LArTPCs.   This paper also
  identifies areas of R\&D for potential improvements in the
  lightguide response.
}


\begin{document}

\section{Introduction}

This paper discusses progress in the development of a lightguide-based
system of light collection in liquid argon, primarily for use in a LArTPC experiment 
such as LAr1~\cite{LAr1} or LBNE~\cite{LBNE}. 
Light collection systems exploit
the fact that charged particles traversing liquid argon produce copious amounts
of ultraviolet scintillation light. The scintillation photon yield is tens of thousands
per MeV of energy deposited~\cite{gastler}, with depending on electric field, local
ionization density and impurity concentrations at the
parts-per-billion level. 

The light has a wavelength of 128 nm and is produced via two distinct scintillation pathways
with different time constants: a prompt component with lifetime of
$\tau= 6$ ns and a
slow component with $\tau = 1500$ ns~\cite{ICARUS}. An intermediate
component with
$\tau = 40$ ns has also been reported by some groups~\cite{WArP}. 
The relative normalization of early to late scintillation
light depends upon the ionization density in the argon, and 
has been utilized as a particle identification variable in some dark matter
searches~\cite{DEAP}.

This paper benchmarks improvements to a recently reported lightguide
detector design for light collection in LArTPCs~\cite{lgnim}. 
It also provides information that is useful for developing monte carlo simulations of
lightguides, for development of future projects. 
In Sec.~\ref{sec:motivation}, we begin with a
brief discussion of the motivation for light collection.  
Then, Sec.~\ref{sec:formula} provide the recipes for two coatings that have been studied.  
In Sec.~\ref{sec:bar}, 
we describe characteristics of the acrylic bars used in this study.
Next, in Sec.~\ref{sec:apparatus}, 
we provide a description of the apparatus used to test the lightguide response, 
followed by a discussion of the waveform analysis (Sec.~\ref{sec:analysis}). 
Sec.~\ref{sec:castacryl} presents the study of the lightguide response, and lastly, 
Sec.~\ref{sec:conclusion} summarizes our results.   
Throughout the discussion, we identify points
where further R\&D are likely to produce substantial advances.

\section{Light collection in LArTPCs\label{sec:motivation}}

\subsection{Liquid argon time projection chamber  (LArTPC)}

In development of LArTPC detectors \cite{uB, argoneut,ICARUS}, most of the attention has
focused on collection of charge to reconstruct tracks to very high precision.    
When charged particles traverse the detector,  ionization electrons are liberated
from the argon atoms by particles traversing a body of liquid argon. 
These electrons
are drifted by an electric field and measured with crossed wire planes to form
a  3D image of the charge deposits left along the tracks of neutrino interaction
products.  The coordinates of the charge deposits perpendicular to the wire
planes is obtained from the drift time for the ionization electrons to reach the
wires.   To obtain the absolute drift time, and hence the perpendicular
coordinate of the event in the detector volume, the time
that the primary interaction occurred, $t_0$, is used.  

\subsection{Motivation}

Recently, attention has turned to establishing light collection
systems.
By collecting and measuring the argon scintillation light we can record
the time structure of the event with few-nanosecond precision and determine the
real $t_0$ of the event.  This provides a method for
establishing $t_0$ in cases where beam timing cannot be used.
This is a much faster and more broadly applicable method than
using the attenuation of charge of known MIP particles to establish
the drift start-location.    

We must know $t_0$ and hence the absolute drift distance
for several important reasons.
Firstly, liquid argon TPCs are commonly run in pulsed neutrino beams,
with beam spills on the order of a microsecond. Surface based TPCs such
as MicroBooNE and some LBNE options expect to be bombarded with a high
rate of cosmic rays and secondary cosmogenic particles such as spallation neutrons
which can mimic neutrino interactions. 
It is therefore vital to determine the
interaction time with microsecond precision to veto the events which occur outside
of the beam window. Once it has been determined that an interaction occurred
within the beam window, the TPC can be triggered and read out. 
The recorded TPC image will in general contain several interactions: 
some corresponding to cosmogenic particles 
and one being the neutrino event we are seeking.
Determining which interaction corresponds to an incident neutrino involves
utilizing geometrical information provided by the optical systems in combination
with information about the event topology from the TPC.
During charge drift, diffusion and recombination of the ionization charge
will occur. Hence there are drift distance dependent corrections which must
be made to correctly measure the $dE/dx$ of a track. Assuming neutrino
events from the beam can be identified, $t_0$ is known from the beam structure and the necessary
corrections can be applied. However, for physics searches which involve events
without a known time of arrival, such as proton decay and supernova neutrino
searches, $t_0$ must be measured by the optical system in order to apply the
required corrections and make an accurate track energy measurement. 

Triggering on the information from the optical system has other practical
benefits from a technical point of view. 
A typical TPC neutrino detector will have a tremendous amount of channels, 
and forming trigger logic on
such a large set is a complicated procedure. 
In contrast, a PMT based optical
system can achieve coverage of the volume with tens of elements, and forming
a trigger becomes more straightforward.
Finally, there are possible applications of the optical information for particle
ID by pulse shape discrimination. The fast to slow scintillation yield ratio
can reveal information about the local ionization of a track, which may be
particularly helpful for performing particle ID on very short tracks where a
TPC based dE/dx determination is either unreliable or impossible.

\subsection{Light Collection Using PMTs}

Cryogenic photomultiplier tubes (PMTs),  
which have a photocathode with platinum undercoating, 
can be used for light collection at LAr temperatures (87 K).  
However  128~nm scintillation light cannot penetrate any glass windows. 
Also, typical bi-alkali PMTs are only sensitive to visible light, 
not the 128~nm scintillation light from LAr. 
Therefore, in a PMT-based system, the light must be
shifted to longer wavelengths.

The favored solution in LAr detectors for shifting
the 128 nm light has been to use a tetraphenyl-butadiene (TPB) layer
between the detector and the PMTs.  This fluorescent wavelength-shifter
absorbs in the UV and emits in the visible with a peak at $425\pm20$~nm~\cite{gehman}, 
which is a favorable wavelength for detection by bi-alkali PMTs.  
Many detectors have used PMTs directly coated with TPB on windows, 
applied as either an evaporative coating (ICARUS~\cite{ICARUS}) 
or embedded within a polystyrene (PS) matrix (
WArP~\cite{WArP}).
The MicroBooNE design separates the coating from the PMT by applying
a TPB-PS mixture to an acrylic plate positioned directly in
front of the PMT.    

Generally the light collection systems of the large active volume detectors 
have favored the use of large PMTs, 
such as the Hamamatsu 5912-02mod 8-inch PMT used in MicroBooNE~\cite{uB}, 
which are sparsely distributed for 
economic reasons.    These tubes are located in the field-free region
of the detectors, typically behind of the TPC anode wire planes. 

\subsection{Lightguides as an Alternative System}

In Ref.~\cite{lgnim}, we presented the first detection of
scintillation light in liquid argon using a lightguide system.  That
paper discussed how  coated acrylic bars, arranged side-by-side as a
paddle, and bent to guide light adiabatically to a single 2 inch
cryogenic PMT, could provide a flat-profile light detection system that
could potentially be inserted into dead regions between LArTPC wire
planes.  This potentially could provide an economical light collection 
system to collect light, if the design is sufficiently efficient.

The lightguides utilized a TPB-based coating with
an index of refraction that was chosen to match acrylic bars.  Acrylic
was chosen as the substrate because it is resilient to cryogenic
cycling and can be easily bent to the required form.  Some of the
visible light that is emitted when UV photons hit the TPB coating will
undergo total internal reflection because the acrylic has an index of
refraction for blue light  ($n=1.49$) that is higher than that of liquid  
argon ($n=1.23$)~\cite{index}.

The reported guides used a polystyrene-TPB coating, as suggested in
Ref.~\cite{McKinsey}, mixed in a 3:1 mass ratio.  The mixture was
dissolved in toluene for application as a liquid.  This was the
highest mass ratio that could be achieved without the TPB crystallizing
on the surface of the guide.     Crystallization must be avoided in
lightguide coatings 
because the white crystals cause absorption and scattering of
visible light as it is reflected along the bars, reducing the attenuation 
length.  With this design, we were able to
demonstrate light collection, albeit with fairly low efficiency.  

\section{Lightguide Coatings\label{sec:formula}} 

We use two coatings in this study.  The first was the same coating as
used the study of Ref.~\cite{lgnim}, called PS25\%.   The second
coating is a new recipe using UVT acrylic, called UVT33\%.   We will
show below that the responses of the two coatings are nearly
equivalent, and so either can be used in future detector designs.

\subsection{PS25\% Coating} 

This coating consists of:
\begin{itemize}
\item 1:3 TPB:polystyrene ratio by mass
\item 50 ml of toluene for every 1 g of polystyrene
\end{itemize}

This is the coating used for our first lightguides \cite{lgnim}. 

\subsection{UVT33\% Coating}

This coating consists of:
\begin{itemize}
\item 1:2 TPB:acrylic ratio by mass
\item 50 ml of toluene for every 1 g of acrylic
\item 1:5 ethyl alcohol: toluene ratio in volume
\end{itemize}

This coating uses UVT acrylic and produces a clear, high
quality coating for use on waveguides. The TPB and UVT pellets 
are first dissolved in the toluene, then the ethanol is added.   
The amount of plastic chosen for each coating saturates the toluene solution, and the amount of TPB chosen saturates the plastic.
The coating is 
applied to the acrylic guide in a single brush stroke. 
From the amount of solution used to coat, 
we estimate approximately $5.5\times 10^{-5}$ g/cm$^2$ 
of TPB is deposited on the surface.  While this sounds like a small amount, it is only a factor of 4 less than an evaporatively coated plate \cite{gehman}.  Applying more solution has limited value, as the attenuation length of 128 nm light is very short in acrylic and PS, meaning only the top portion of the coating is active.
The thickness cannot be related to the performance of the coated plate easily, 
because other factors, including coating method and surface condition, 
are equally important.   An important goal of this paper is to study the effect of local variations in the TPB concentration. See Section \ref{results}.

Addition of ethyl alcohol improves attenuation length because  
acrylic dissolved directly in toluene produces a rougher
coating, more likely to scatter the light in the guides.   
Ethyl alcohol thins the coating as well as smoothing it. 
However, addition of ethyl alcohol may introduce negative effects, such as self-absorption. 
As we show later (Figure~\ref{fig:SpectrometerAttachments}), 
the guided light at the TPB emission peak (436~nm) 
shows fairly uniform response against the incident spectrum. 
However, a few drops of ethyl alcohol introduces slight overall reduction ($<$10\%) 
and a small dip at 270~nm.   
We have varied the fraction of alcohol to find the optimum ratio to Toluene. 

\subsection{Tests of Coatings in a Vacuum Monochromator}

\begin{figure}
\centering
\includegraphics[width=0.50\textwidth]{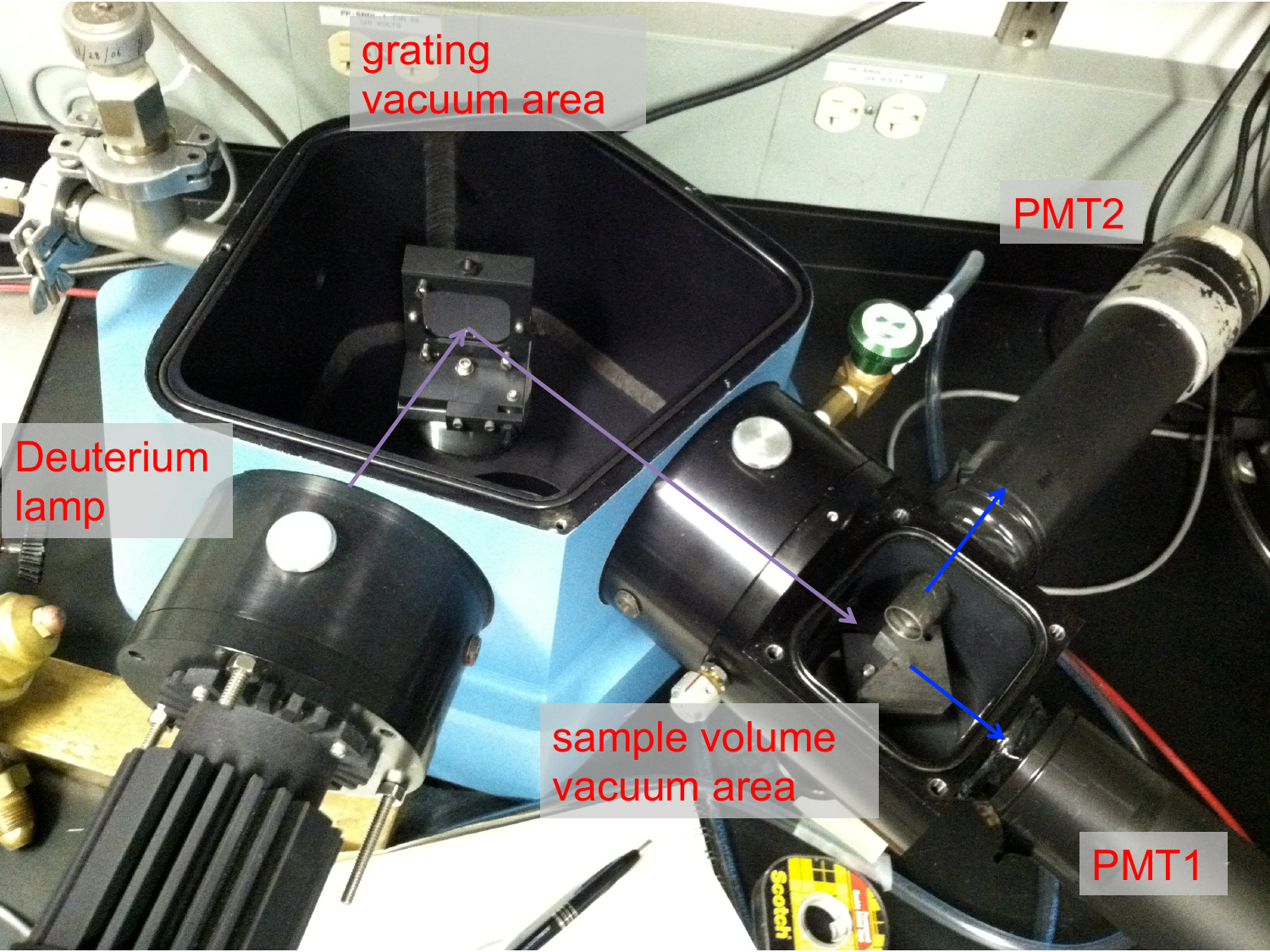}
\caption{Setup for the vacuum monochromator measurement. 
Light from the deuterium lamp enters a vacuum grating area 
where a specific wavelength can be selected by the grating to impinge on a sample. 
The emission from the sample can be observed by 2 PMTs located at built-in windows to the sample chamber.  PMT 1 is used for this study.
\label{vacuumspec}}
\end{figure}

Figure~\ref{vacuumspec} shows the setup of the vacuum  monochromator test. 
Tests were performed in a McPherson 234 vacuum monochromator using 
a McPherson model 632 UV
Deuterium Lamp.  
Measurements were taken at a pressure of 
11~$\pm$~4~mTorr at room temperature, 
though results at 215~nm and 250~nm are consistent with measurements at atmospheric pressure in the same setup. 
A PMT located outside of the vacuum region is used to see the forward emission 
of the sample plate.  PMT 1 in figure \ref{vacuumspec} was used for this measurement.

\begin{figure}
\centering
\includegraphics[width=0.65\textwidth]{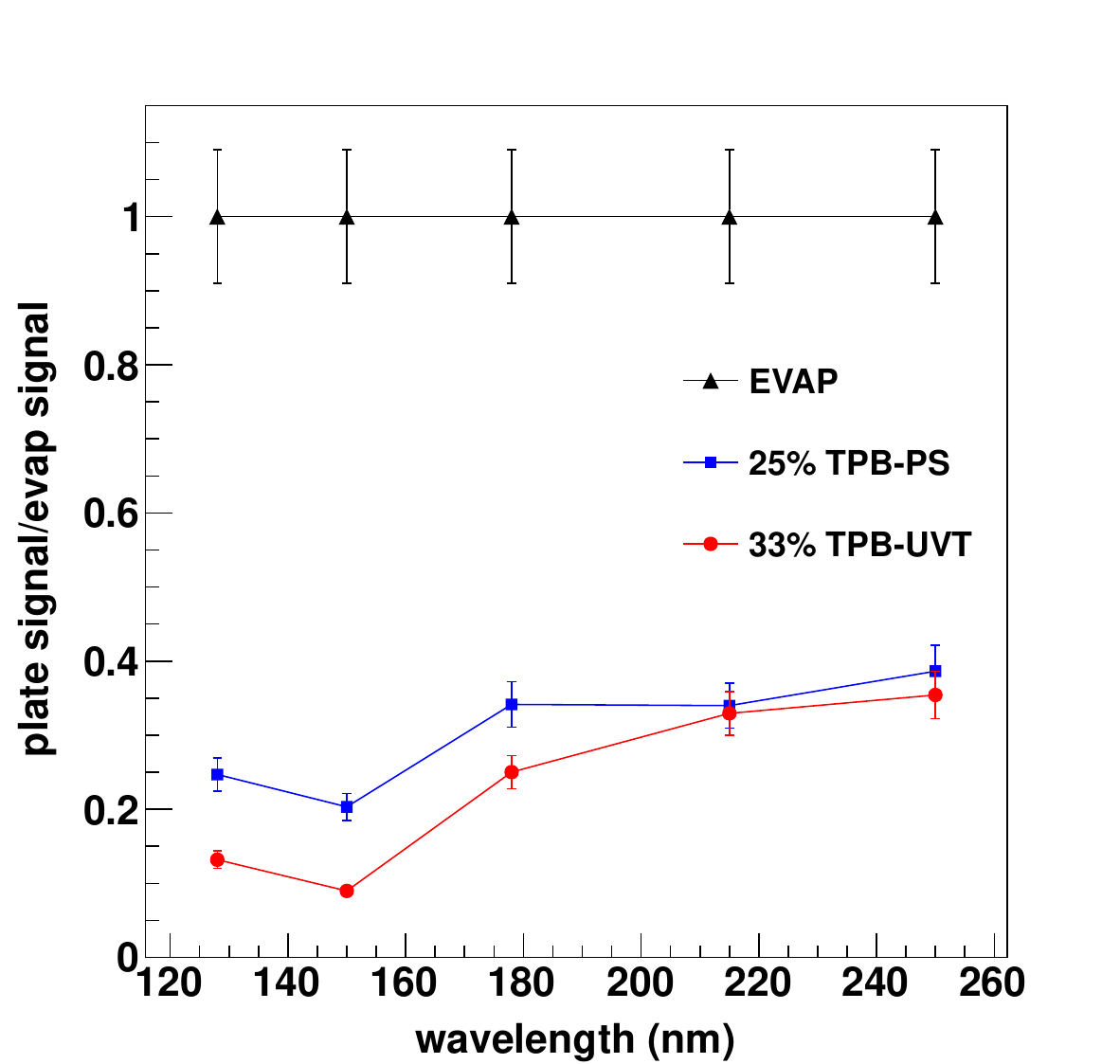}
\caption{Coating efficiency as measured in a vacuum monochromator 
as a function of wavelength.  
Samples are normalized to an evaporatively coated reference sample 
in order to calibrate out the spectrum of the Deuterium lamp. 
Error bars on the UVT and PS points only take into account 
the errors associated with these measurements, 
allowing for comparison between the two types of 
samples with each other, 
while error bars on the evaporative sample illustrate 
additional errors in the efficiency measurement.     
\label{avemono}}
\end{figure}

The data in Figure~\ref{avemono} represent average measurements and 
are normalized with respect to average measurements of evaporatively coated plates.  
The evaporative samples contain pure TPB evaporated onto a plate in vacuum, 
and thus the effects of the different matrices can be compared to the pure TPB response. 
The evaporative samples used in this study have coating thicknesses of 1.87~$\mu$m. 
We expect these to be somewhat similar in performance to the plates used in reference~\cite{gehman}, 
which had a thickness of 1.5~$\mu$m.
The errors on each plate are associated with statistical errors 
including testing different samples of each coating type and 
systematic errors associated with the vacuum monochromator.  
The overall efficiency including the error on the evaporative measurement 
at 128~nm of our two types of coatings relative to 
the previously described evaporatively coated plates are
0.25~$\pm$~0.03 and 0.13~$\pm$~0.02 for 
the PS25\% and UVT33\% coated acrylic respectively. 

Note that this result appears to contradict Sec. 7, where we will present that the UVT coating yields higher efficiencies when applied to light guides than the PS coating.  We believe this to be caused by light being captured in the TPB-PS coating, which has a higher index of refraction (n=1.59) than that of the base acrylic (n=1.49).  The light trapped in the coating is almost immediately lost through surface attenuation due to the thinness of the coating, appearing as a drop in the overall coating efficiency

\subsection{Emission Spectra From Light Guides}

\begin{figure}
\begin{centering}
\includegraphics[trim=0cm 5cm 0cm 0cm, width=0.75\columnwidth]{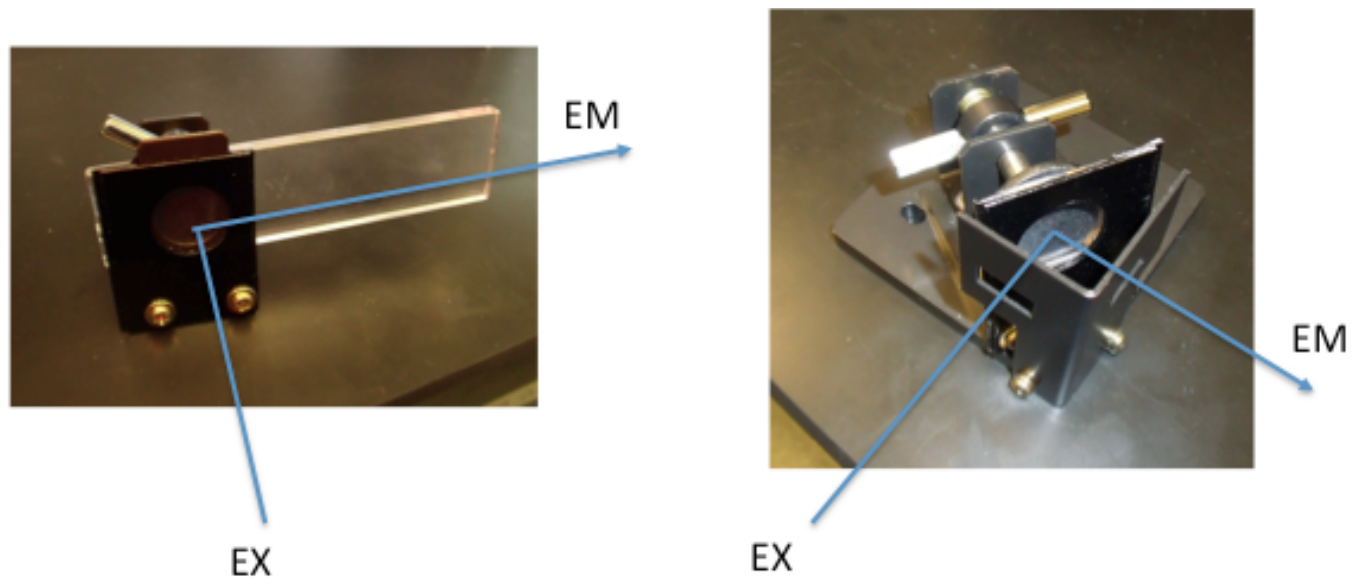}
\par\end{centering}
\caption{Spectrometer attachments used to obtain guided (left) and backward 
(right) emission and excitation spectra for lightguide sections. 
\label{fig:SpectrometerAttachments}}
\end{figure}

The guided emission spectra from several short light guide segments
of up to 10 cm in length were measured using a fluorimeter.
A tunable monochromatic beam, produced using a grating and a xenon
lamp, is normally incident upon the TPB coated surface of the light
guide. The guided spectrum is measured at 90 degrees to the incident
beam through of the end of the light guide using a second tunable
grating and a photomultiplier tube. The spectrometer attachment used to
implement this configuration is shown in figure \ref{fig:SpectrometerAttachments}
(left). 

The lamp spectrum and PMT response are automatically accounted for
by the spectrometer software, and the device is routinely re-calibrated
using two standard samples : a rhodamine dye sample to characterize
the emission grating and PMT response, and a diffuse glass cuvette
to characterize the excitation grating and xenon lamp spectrum.

\begin{figure}
\begin{centering}
\includegraphics[width=0.8\columnwidth]{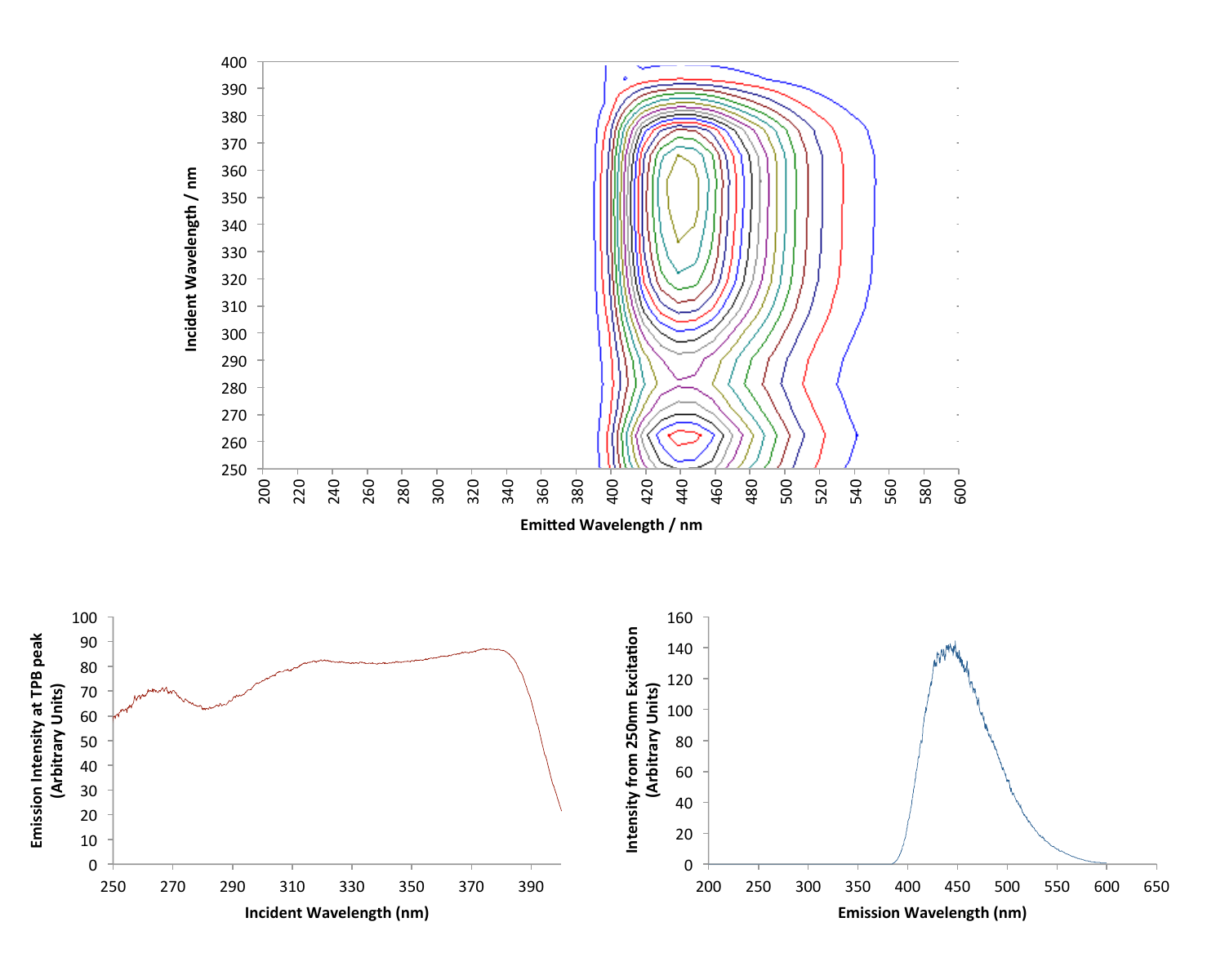}
\par\end{centering}

\caption{Emission and excitation spectra of wavelength shifted and guided light
from a short light guide section with UVT33\% coating. 
The top plot shows the two-dimensional emission excitation spectrum, 
bottom 2 plots are the slices of 436~nm emission (left) and 
250~nm excitation (right). 
 \label{fig:GuidedEmExSpectra}}
\end{figure}

The guided spectrum was measured between 200 and 600 nm for incident
wavelengths between 200 and 700nm. For segments of length 6 cm, 8cm
and 10 cm, no differences in the shape of the emission or excitation
spectra were seen. For incident light above 400 nm, no wavelength shifting
behavior is observed, so we omit this region from the reported plots.
Figure \ref{fig:GuidedEmExSpectra}, top 
shows the two dimensional emission-excitation
spectrum as a contour plot. We also show the emission spectrum at 250nm 
(Fig.~\ref{fig:GuidedEmExSpectra}, bottom left), 
and the wavelength shifting capability at the TPB emission
peak wavelength of 436~nm (Fig.~\ref{fig:GuidedEmExSpectra}, bottom right). 
These one dimensional plots can be interpreted
as a single horizontal and vertical slice from the two dimensional
contour map, respectively.

\begin{figure}
\begin{centering}
\includegraphics[width=0.75\columnwidth]{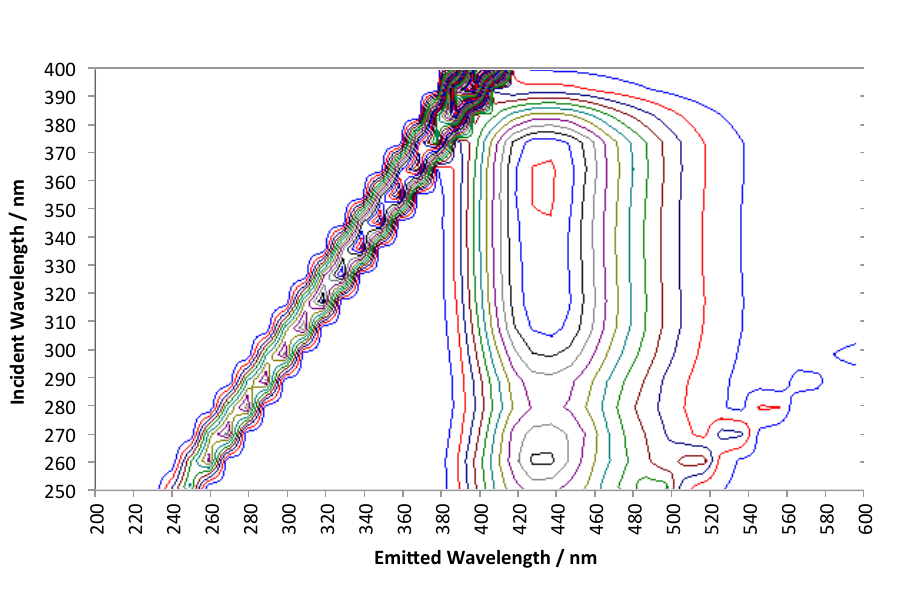}
\par\end{centering}
\caption{The two-dimensional backward emission and excitation spectra for the UVT33\%
and a thickly coated acrylic plate. 
The linear feature at EM=EX is directly reflected incident
light. The second linear feature at lower wavelengths is a spectrometer
alias due to the high intensity of direct light, and should be disregarded.
\label{fig:SurfaceEmExSpectrum}}
\end{figure}

The two dimensional excitation-emission spectrum for a backward emission 
was also measured. In this setup, a short light guide
section is illuminated at 45 degrees to the surface with a monochromatic
beam. The emitted light at 90 degrees to the incident beam is detected.
The spectrometer attachment used to implement this arrangement is
shown in figure \ref{fig:SpectrometerAttachments} (right). 
The two dimensional backward
excitation-emission spectrum for the light guide coating
is shown in figure \ref{fig:SurfaceEmExSpectrum}.

\section{Bars Under Study\label{sec:bar}}

In the discussion below we consider coatings applied to extruded and cast
acrylic bars to form light guides.  Cast acrylic is taken to be the
standard in our study.  Our cast bars are purchased from Altec Plastics \cite{Altec} and
are polished on the ends.   This replaces the extruded acrylic bars
used in Ref.~\cite{lgnim}  purchased from
McMaster-Carr \cite{McMC} that are substantially more economical, however,
have significant imperfections.

\subsection{Attenuation Length Measurements of Uncoated Acrylic Bars \label{attennocoat}}

\begin{figure}[tb]
\centering
\includegraphics[width=0.4\textwidth]{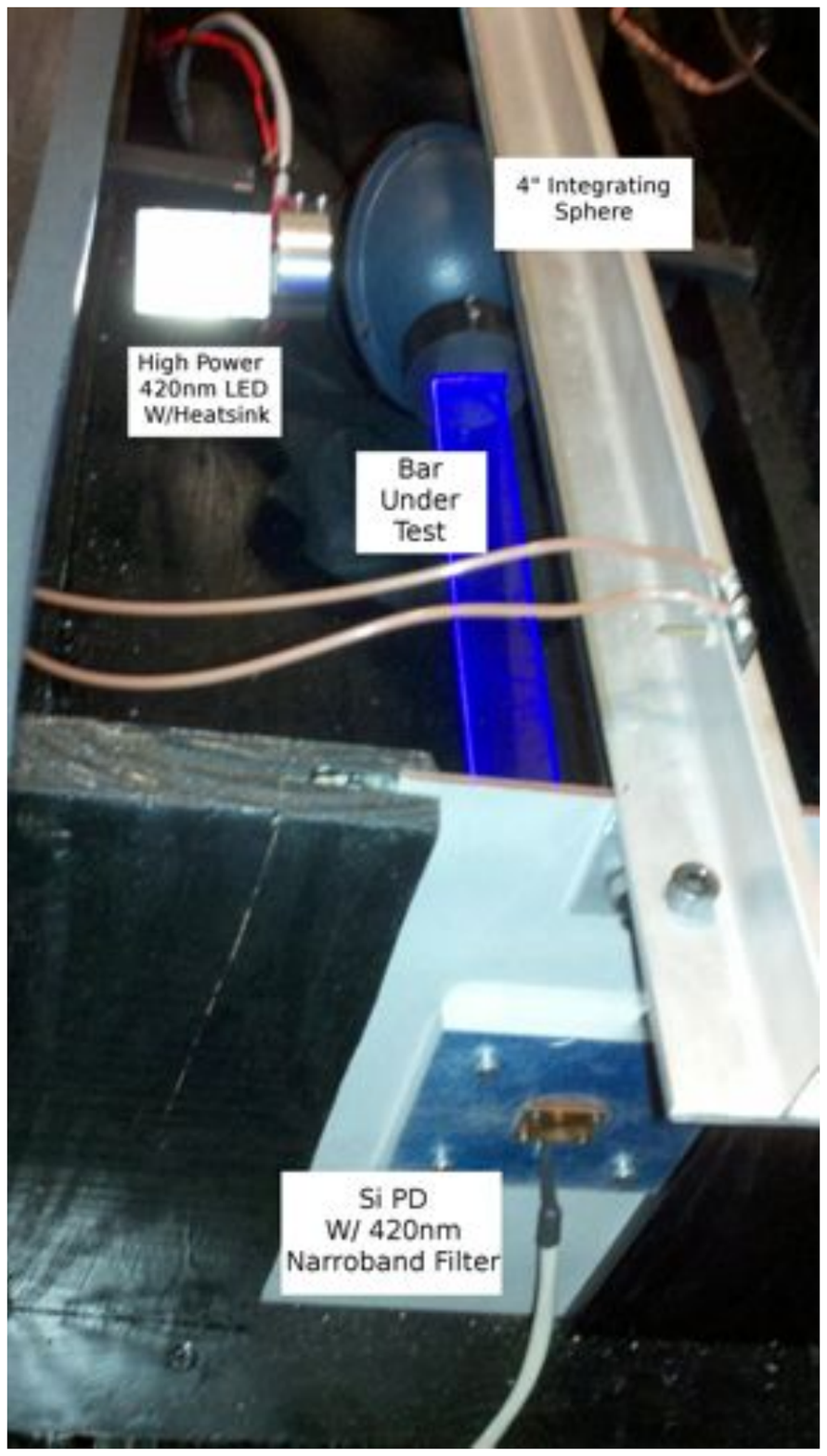}
\caption{Setup used to measure the attenuation length of uncoated extruded acrylic and cast acrylic bars.}
\label{intsphere_setup}
\end{figure}

We measured the attenuation length of uncoated acrylic bars by illuminating 
the end of rods cut to various lengths and recorded the photo-current observed 
by a Si photodiode (PD). 
We used acrylic manufactured in two different ways: a cast UVA acrylic and 
an extruded UVA acrylic. The setup can be seen in figure \ref{intsphere_setup}.

The detector used to measure the light emerging from the end of 
the bars was an OSI Optoelectronics 3~mm Si PD (P/N: OSD15-0). 
In this configuration we are overfilling the detector and therefore 
only sampling a small fraction of the end of the bar. 
In addition, a $420\pm10$~nm narrow band filter was placed in front of 
the detector to restrict the attenuation length measurement to 420~nm. 
The illumination was performed with a high power $420\pm10$~nm LED (FutureLED P/N: FL-LED-440-420) 
driven at a constant current of 187~mA. 
The LED light was input into an integrating sphere to diffuse the light, 
thus illuminating the rod end with a uniform illumination pattern. 
The PD was mounted so that it views only the center of 
the bar through the narrow band filter. 
If the LED light is passed directly down the bar, 
the illumination pattern at the output of the bar is nonuniform and slight misalignments between the LED, 
light guide and PD can lead to systematics 
in the measurement of the attenuation length. 
Uniform illumination from an integrating sphere, 
on the other hand, will reduce this systematic significantly. 
This systematic would be minimized in a LAr detector design that under fills the PMT photo-cathode area.

Two sets of ten bars, one set for each type of acrylic, 
were measured with lengths increasing by 7.62~cm (3~inches) between 7.62~cm and 76.2~cm (30~inches). 
The $0.64 \times 2.54$~cm$^2$ (1/4~inch~$\times$~1~inch) cast UVA acrylic bars, 
purchased from Altec Plastics, 
were cut out of large acrylic 0.64~cm thick sheets. 
The cut edges on the 0.64~$\times$~76.2~cm$^2$
face were polished by Altec Plastics using a polishing machine. 
The quality of measurement of attenuation length depends strongly on how well the sides are polished. 
The  $0.48 \times 2.54$~cm$^2$ (3/16~inch~$\times$~1~inch) 
extruded UVA acrylic bar, purchased from McMaster-Carr, 
didn't require polishing on the 0.48~$\times$~76.2~cm$^2$ face, 
because the bars were extruded at the 2.54~cm width and 
the factory edge was suitable for making this measurement.
In both cases the ends of the bars were polished in 
the lab using a diamond tipped fly cutter after they were cut to length.

\begin{figure}[tb]
\centering
\includegraphics[width=0.7\textwidth]{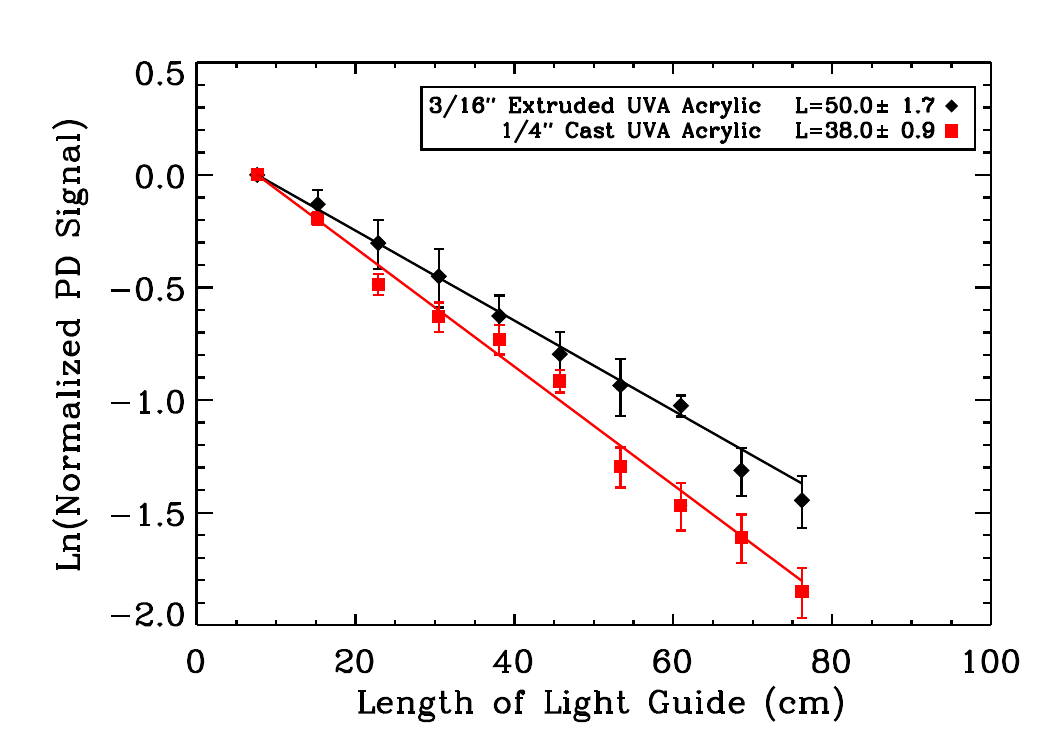}
\caption{Attenuation length measurements of extruded and cast UVA acrylic bars at 420~nm. 
A fit to the data resulted in a measured attenuation length of 50$\pm$2~cm for the extruded UVA acrylic and 38$\pm$1~cm for cast UVA acrylic.}
\label{atteniu}
\end{figure}

Fig.~\ref{atteniu} shows natural log of the normalized photo-current versus 
the bar length for both the extruded and cast UVA acrylic. 
A fit to these data resulted in a measured attenuation length of 
50$\pm$2~cm for the extruded UVA acrylic and 
38$\pm$1~cm for cast UVA acrylic.

\subsection{Preparation and Storage of Coated Bars}

All bars used in the tests below are 60~cm in length and have polished
ends.     The cast bars are delivered with polished ends, while the
ends of the extruded bars are polished in the lab.   
Before coating, bars are cleaned with ethyl alcohol. 
The coatings are applied with one brushstroke using an acid brush,
depositing about $5.5\times 10^{-5}$ g/cm$^2$ of TPB

There is clear evidence that TPB coatings degrade with even modest
exposure to laboratory fluorescent lights and sunlight~\cite{degradation}.   
Our studies show a 30\% loss of response after a single day of exposure
and 80\% degradation after one month.   Therefore the bars are handled
in a laboratory with UV filters installed on the fluorescent lights
and windows.     Furthermore, whenever possible, bars are kept covered 
with light-blocking cloths or are stored in opaque containers.

The previous study of lightguides predated the demonstration of the 
detrimental effect of UV light.  Therefore, the precautions described
above were not taken with the lightguides discussed in 
Ref.~\cite{lgnim} .   We think that this is the primary explanation for
why the outputs of the PS25\% lightguides presented in this paper are nearly a factor
of three higher than that reported in Ref.~\cite{lgnim}.  

There is some evidence that TPB coatings degrade by about 10\% due to 
exposure to humidity in the laboratory~\cite{WarpH2O}.
This can be mitigated
by desiccating the bars.   
The effect of humidity is relatively small
and the evidence for improvement modest at best.  Nevertheless, we store the 
bars used in these studies in containers with desiccant packets.

\section{Apparatus for LAr Tests\label{sec:apparatus}}

The lightguide test stand was described in detail in
Ref.~\cite{lgnim}.  The test stand is constructed from 
an open-top glass dewar which is 100 cm tall and 
14 cm inner diameter into which a holder containing the 
lightguide is 
inserted.   The holder has a 
7725-mod Hamamatsu, 10-stage PMT~\cite{Hamamatsu}
with a custom cryogenic base attached at the bottom.   
One PMT was used for all tests reported here.    In the LAr, the 
PMT floats up against the lightguide which is fixed in the holder,
making a good optical connection.  An improvement to the holder 
of Ref.~\cite{lgnim} guides the PMT so that the relative 
position between PMT face and bar is reproducible.

Lightguides are tested with 5.3~MeV $\alpha$
particles produced from a $^{210}$Po source mounted in a plastic disk~\cite{Unitednuclear}.  
The source is electroplated onto foil that is
recessed into a 3mm ``well,'' of the plastic disk.  In an improvement
over Ref.~\cite{lgnim}, the disk is held in a holder with a 5.1 mm
diameter hole, leading to a well of 4.8 mm depth.  
The $\alpha$s
emitted into the well traverse $\sim$50~$\mu$m in LAr. Scintillation
light is then isotropically emitted.   
The well occludes most of the
light; the solid angle acceptance for light at the bar is 7\%.

\begin{figure}[tb]
\centering
\includegraphics[width=0.75\textwidth]{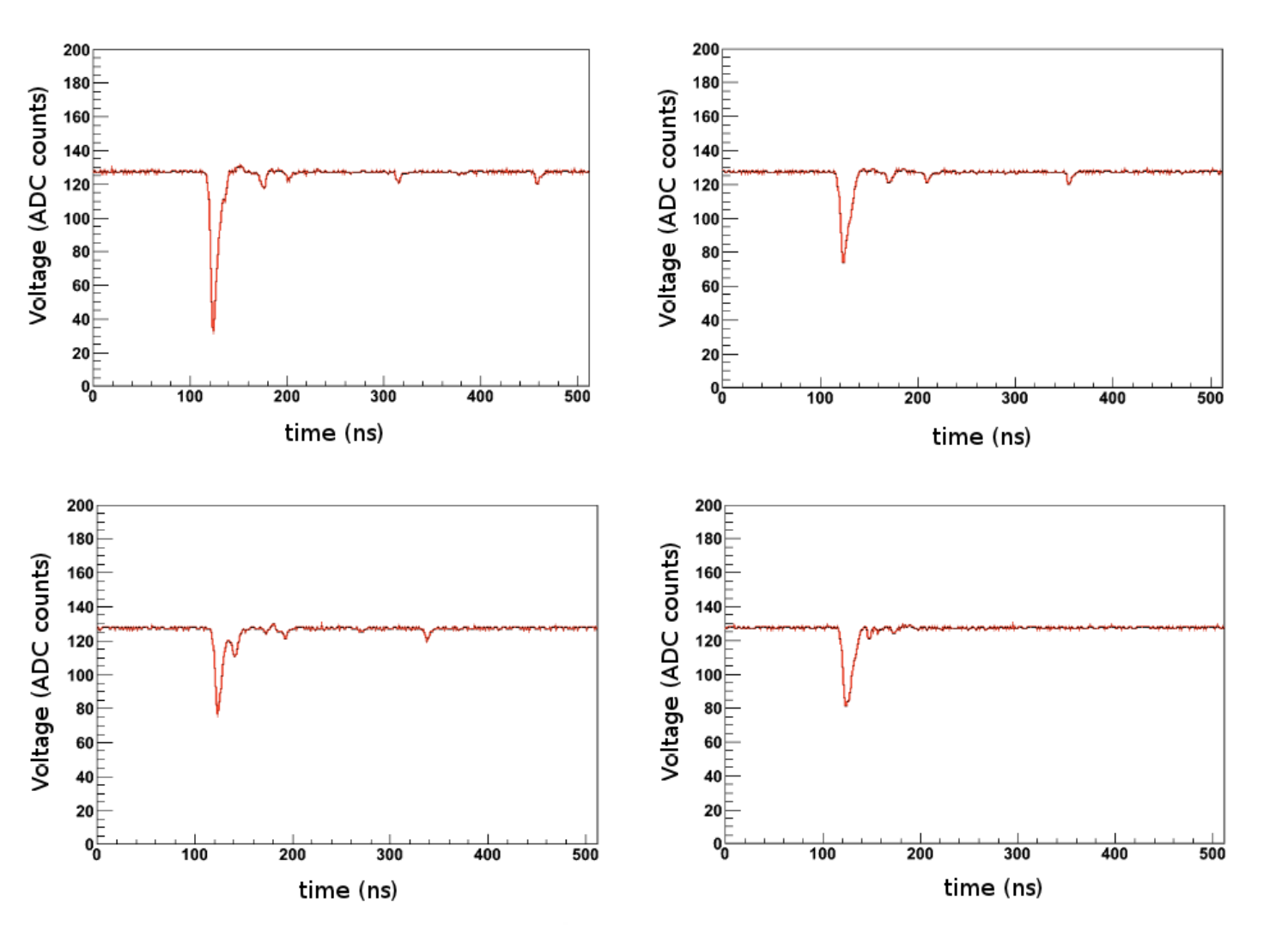}
\caption{Four example events showing 
readout of waveform digitizer on a scale of $\pm$200~mV/256~ADC counts. 
The scintillation light produced by the
$\alpha$ has an early and late component.
The triggering large pulse is primarily early light.  The following
small pulses are from single photons due to the late component.  
\label{pulses}}
\end{figure}

The holder is immersed in LAr for 30~minutes before data is taken.  
Readout of the PMT is performed using an Alazar Tech ATS9870
digitizer. The Waveform is recorded on a scale of 200~mV/128~ADC counts and a trigger
is produced by a negative pulse with an amplitude that
exceeds 17~ADC counts, corresponding to a $-27$~mV threshold.  
When a trigger is produced, 128~pre-trigger samples and 384~post-trigger samples 
are recorded at a sampling rate of 1 giga-sample
per second, leading to a total recorded profile of width 0.512~$\mu$s.

Triggers from the $\alpha$ source occurred at a rate of about 300 to 400~Hz, 
which was consistent with expectation given the short lifetime of the
polonium.   Runs were taken with no $\alpha$ source in order to
measure the cosmic ray rate, which was found to be about 8 Hz.  
The dark rate, measured in runs with no light guide with the above threshold, was $<$1~Hz.

Our studies use four batches of industrial grade LAr.   
Industrial grade LAr
is CGA certified to contain $<$20~ppm
nitrogen and $<5$ ppm~oxygen~\cite{CGA}.  
Running with ``ultra-pure grade'' which has
has $<$5~ppm nitrogen and $<1$ ppm oxygen~\cite{CGA}, was
also an option, however we found no difference in the results
of the studies below with the industrial versus pure LAr.
This surprising result may arise for two reasons.  First,
the process of filling the open dewar system may lead to
roughly equal contamination levels of the ultra-pure argon and
the industrial argon.   Second, 
the studies reported here focus on the quantity of early light, but
not the quantity of late light.  Impurities most dramatically affect 
the quantity of late light~\cite{WARPN,WARPO}.  
With this said, further investigation with purified LAr is warranted.  

In order to limit contamination by air when filling, 
the test stand is first filled with argon gas.  The LAr is then poured
through the gas into the dewar.    We create the initial gas layer in
the dewar by half-filling the warm 
dewar and allowing this LAr to evaporate.

\section{Analysis of the Waveforms\label{sec:analysis}}

Fig.~\ref{pulses} shows the waveforms of some example events
from runs with the $\alpha$ source.  One
expects to see an initial peak  corresponding to many photons
followed by pulses from late light that correspond to a single UV
photon hitting the bar distributed in time.       The lifetime of the 
early light is 6 ns, and so one expects 95\% of the photons to be 
produced within the first 18~ns of the pulse.    In fact, the pulses
appear somewhat wider due to fact that some late light also
populates the initial peak, although this component is  
highly quenched due to the impurities in the LAr~\cite{WARPN,WARPO}.
As a result,  the initial pulse is predominantly early light and we will refer to
the initial pulse as early light in the discussion below.

\subsection{The Pulse-finding Algorithm and Variables}

We apply a pulse-finding algorithm that identifies pulses and 
records the maximum pulse-height from the
baseline and the integrated charge of each pulse.   
The trigger is defined as when the signal drops to 17 or more ADC counts below the baseline. 
The pulse is then integrated from 30~ns before the trigger to 120~ns after the trigger. 
This defines total charge, $Q_{tot}$, which is our primary observable.
The long integration period accommodates large initial pulses from cosmic rays 
as well as the smaller initial  pulses from the $\alpha$ source.  
The pulse-height, $Q_{max}$, is defined as the difference between the baseline and 
the  minimum ADC count during the integration period.   
The baseline is recalculated using a 10~ns time window before each pulse, 
for both early and late light pulses.  
The start point for late light pulses is defined 
as when the signal drops to 2 or more ADC counts below the baseline.   

\subsection{The $Q_{tot}$ Distributions of Early Light}

\begin{figure}[tb]
\centering
\includegraphics[width=0.85\textwidth]{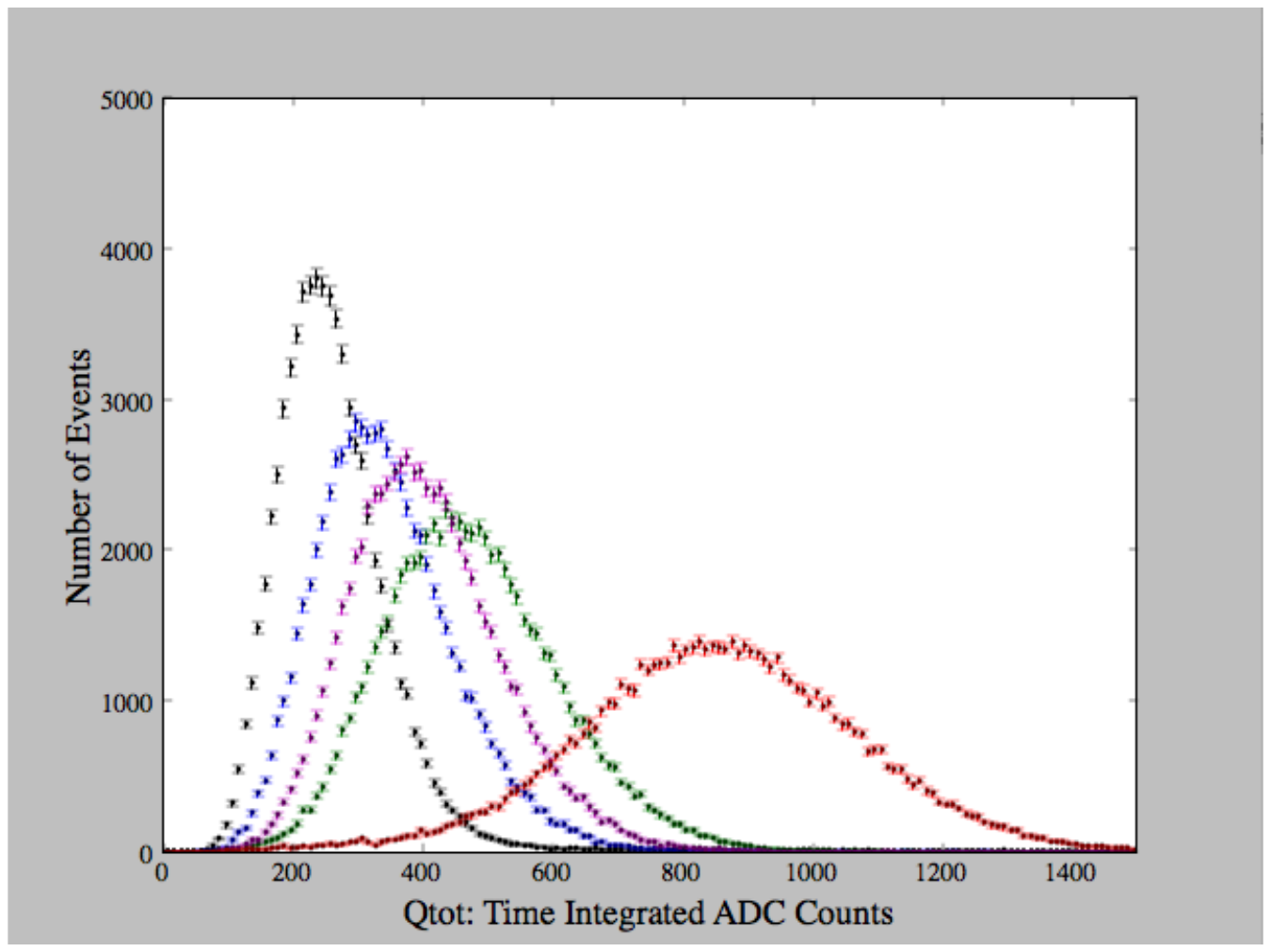}
\caption{(color online). 
A variety of example $Q_{tot}$  from cast acrylic bars with
  acrylic coatings with the source at 10~cm (red, widest distribution), 20~cm (green),
 30~cm (magenta), 40~cm (blue) and 50~cm (black, narrowest distribution).   The data are from
 four bars and four batches of LAr taken during a period of 60~days.    
  All distributions have $>70,000$~events.  
 \label{Qtot}}
\end{figure}

After corrections for attenuation along the lightguide, the $Q_{tot}$
distribution of the early pulses can be correlated to the energy 
of the $\alpha$ in the event.
Examples of $Q_{tot}$ distributions for the early light 
are shown in Fig.~\ref{Qtot}.      
In the analysis below, we will study variations in these distributions
as a function of position and batch of LAr.   Thus, for illustrative
purposes, we provide examples of distributions with the source
at  10 cm (red), 20 cm (green),  30 cm (magenta), 40 cm (blue) and 
50 cm (black) for various lightguides and LAr batches. 
In the analyses below, the  $Q_{tot}$ distributions are taken from runs of 
$>70,000$ events.

Using $Q_{tot}$ is a departure from the study in
Ref.~\cite{lgnim},  which employed pulse-heights, $Q_{max}$.  However,
with the improved efficiency for the lightguides, we find that
$Q_{tot}$ is a better representation of the number of photons in an
event.   The issue arises when the photon arrivals are distributed
over  long time periods, leading to multiple peaks.
An example of such a distortion is seen in the bottom left of
Fig~\ref{pulses}.   This problem was addressed by the ``multipeak analysis''
of Ref.~\cite{lgnim}, however employing $Q_{tot}$ is a simpler
and more accurate solution.

\subsection{Late Light:  A Single Photon Sample }

As can be seen from Fig.~\ref{pulses}, the late light is sparse,
however  the pulses are well-formed and uniform.   The late light is 
particularly valuable, because it allows measurement of the $Q_{tot}$
distribution for one and only one photon arriving at the PMT. 
Therefore, we utilize the late light to calibrate our system.     

Our late light sample is acquired in the range $> 400$ ns after the
initial pulse producing the trigger.   This is sufficiently late in
time that considering the yield and time constant of the light, 
we can be assured that the pulses which arrive at the
lightguide correspond to only one UV photon.     
It has been shown that, on average, 1.3 visible photons are 
emitted from an evaporative TPB coating per one incident UV photon \cite{gehman}; 
which is to say that occasionally
TPB will produce multiple photons rather than one
However, the acceptance of the lightguides, is only 5\%, so there is a negligible
probability of multiple photons arriving at the PMT.   

Most methods of calibration, such as using low intensity LEDs, 
involve a Poisson distribution of photons arriving at the 
PMT  which is then used to find the 1 PE response.   
In contrast, this calibration method is assured to sample exactly one photon hitting the PMT.    
As a result, one expects the $Q_{tot}$ distribution of the late light
will simply reflect the statistics of the early stages of the dynode chain,
which is expected to be 
represented by a Gaussian to a good approximation \cite{PMTresponsefunction}.
As expected, the peak position varies with PMT voltage.  However, for a given PMT, 
set at a specific voltage, the peak position is always located at 40 counts$\times$ns,
regardless of the lightguide being tested.  
Also, we find that, for a specific lightguide-and-source set up, 
if the PMT high voltage is always adjusted such that the 1 PE response peak is at 40
counts$\times$ns, then the prompt light response is
reproducible. 

\begin{figure}[tb]
\centering
\includegraphics[width=0.85\textwidth]{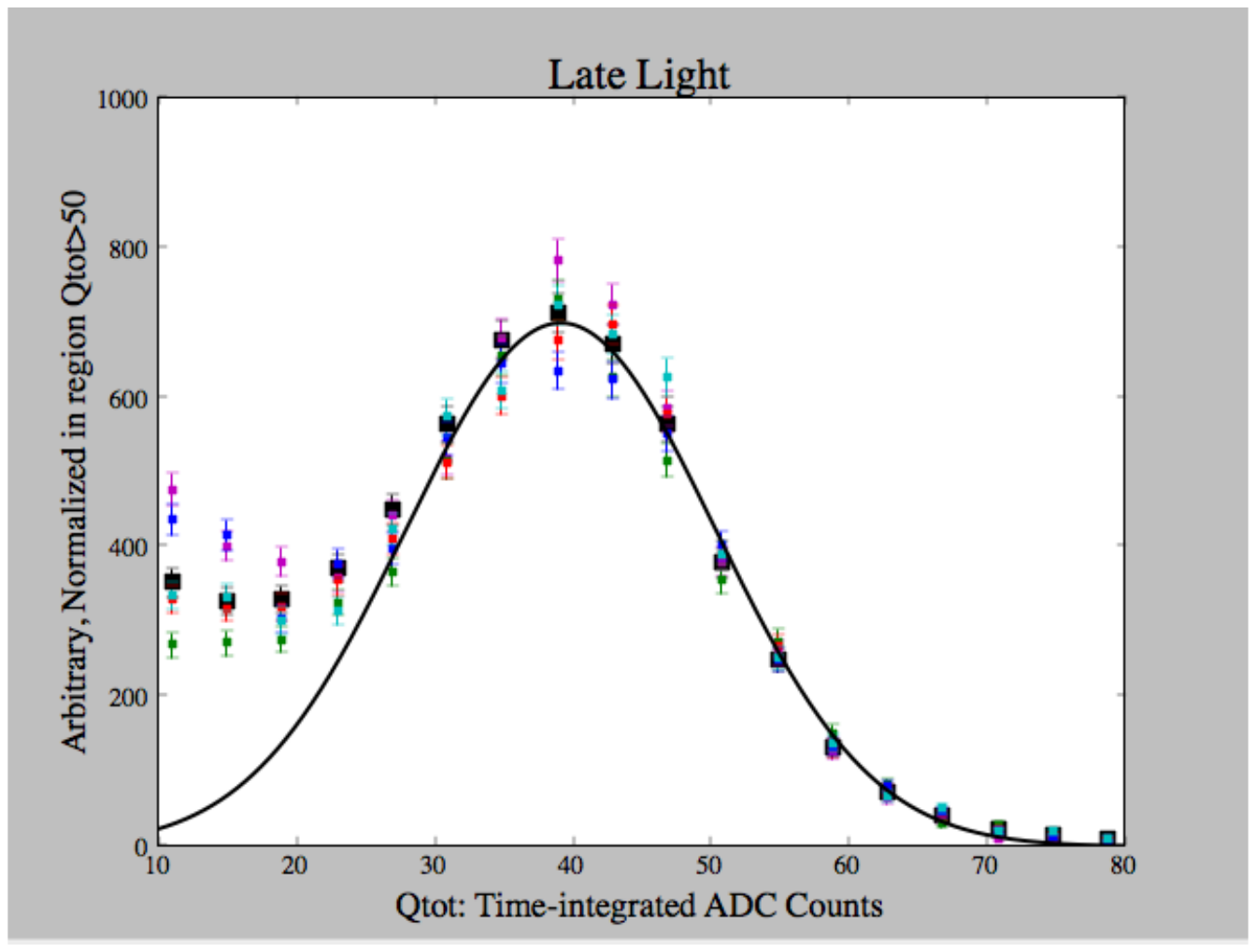}
\caption{(color online). 
Colored symbols: Five example late light examples sampling all of the batches of LAr
  and bars and spanning the typical range of late light
  distributions.  Samples were normalized in the range $Q_{tot}>50$
  counts$\times$ns.  Black squares:  mean of ten late light files,
  after normalization in the $Q_{tot}>50$  counts$\times$ns range.
  The Gaussian is a fit to $Q_{tot}>30$ counts$\times$ns.\label{overlaylate}
}
\end{figure}

We demonstrate this in Fig.~\ref{overlaylate}, 
where we overlay the late light distributions from five example runs. 
The examples cover four bars and four LAr batches and 
are typical of all the late light data sets.  
To allow comparison, the distributions are normalized in the range
$Q_{tot}>50$  counts$\times$ns, where background is expected to be low.   
One sees that the peaks of the late light are in good agreement,
but the $Q_{tot}<30$ counts$\times$ns can vary 
due to the relative strength of the background under the different environment. 
However this does not affect the location of the 1~PE peak, 
as Figure~\ref{overlaylate} show. 
The source of background is under investigation. 
The black squares show the mean of ten late light distributions.

Using the expectation for PMTs with large ($>4$)  secondary electron emission at
the first dynode and high collection efficiency by
the first few dynodes, we find 
a Gaussian fit to $Q_{tot}>30$ counts$\times$ns yields a single PE peak at
39.3$\pm$1.0 counts$\times$ns. 
Thus, in the analysis that follows,  
the conversion from $Q_{tot}$ to observed PE is obtained by dividing the 
results by 39.3 counts$\times$ns/PE.     
This was the same technique as was employed in Ref.~\cite{lgnim}.

\section{Cast Acrylic Bars with Acrylic Coatings\label{sec:castacryl}}

The following tests are performed on cast acrylic light guides with UV33\%
coatings brushed onto cast acrylic bars.

\subsection{Results of Tests in LAr \label{results}}

\begin{figure}[tb]
\centering
\includegraphics[width=0.35\textwidth]{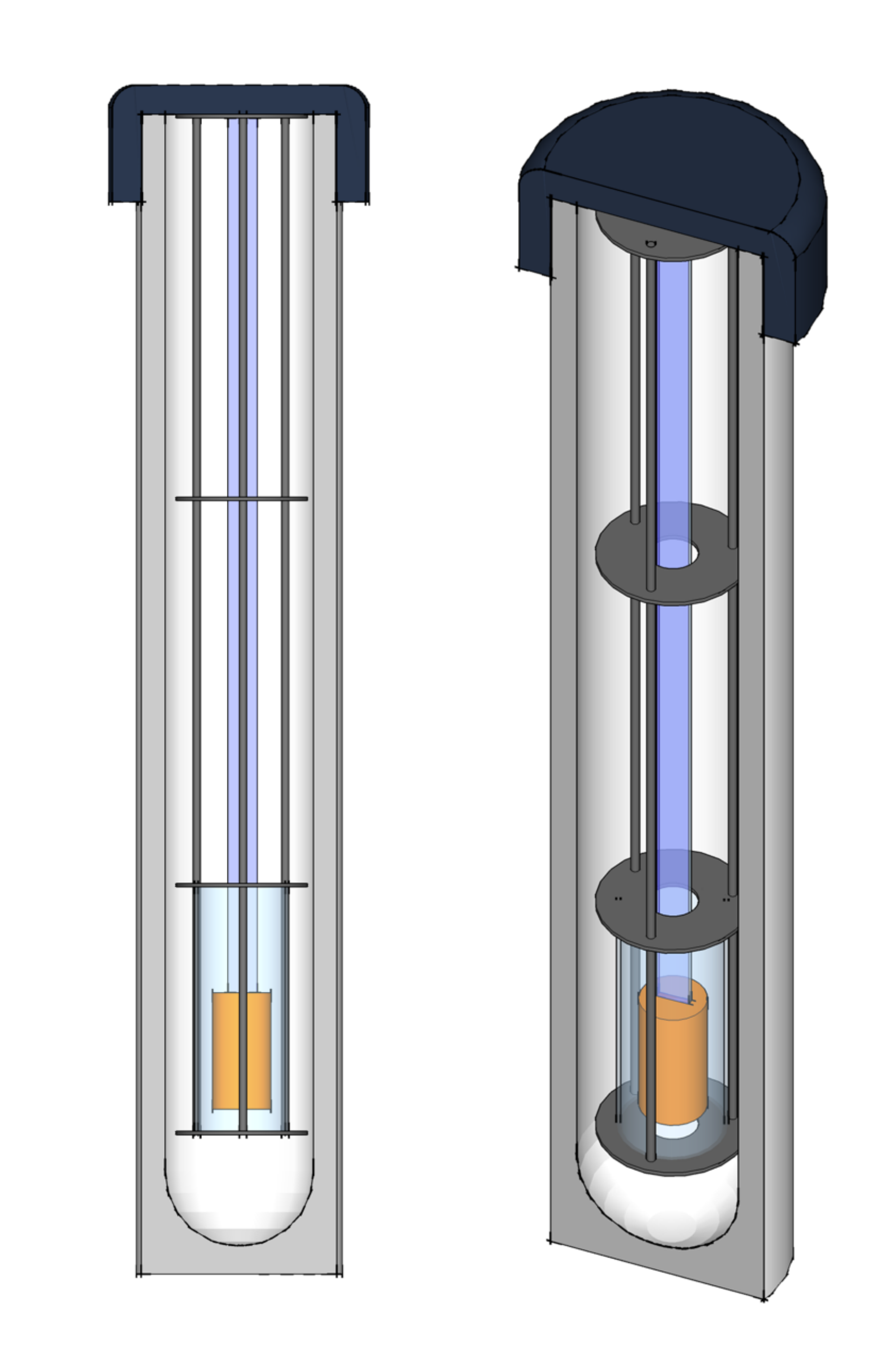}
\caption{
Two views of the test setup in LAr.  The alpha source is placed at various locations along the bar for the described tests.  \label{larsetup}
}
\end{figure}

\begin{figure}[tb]
\centering
\includegraphics[width=0.95\textwidth]{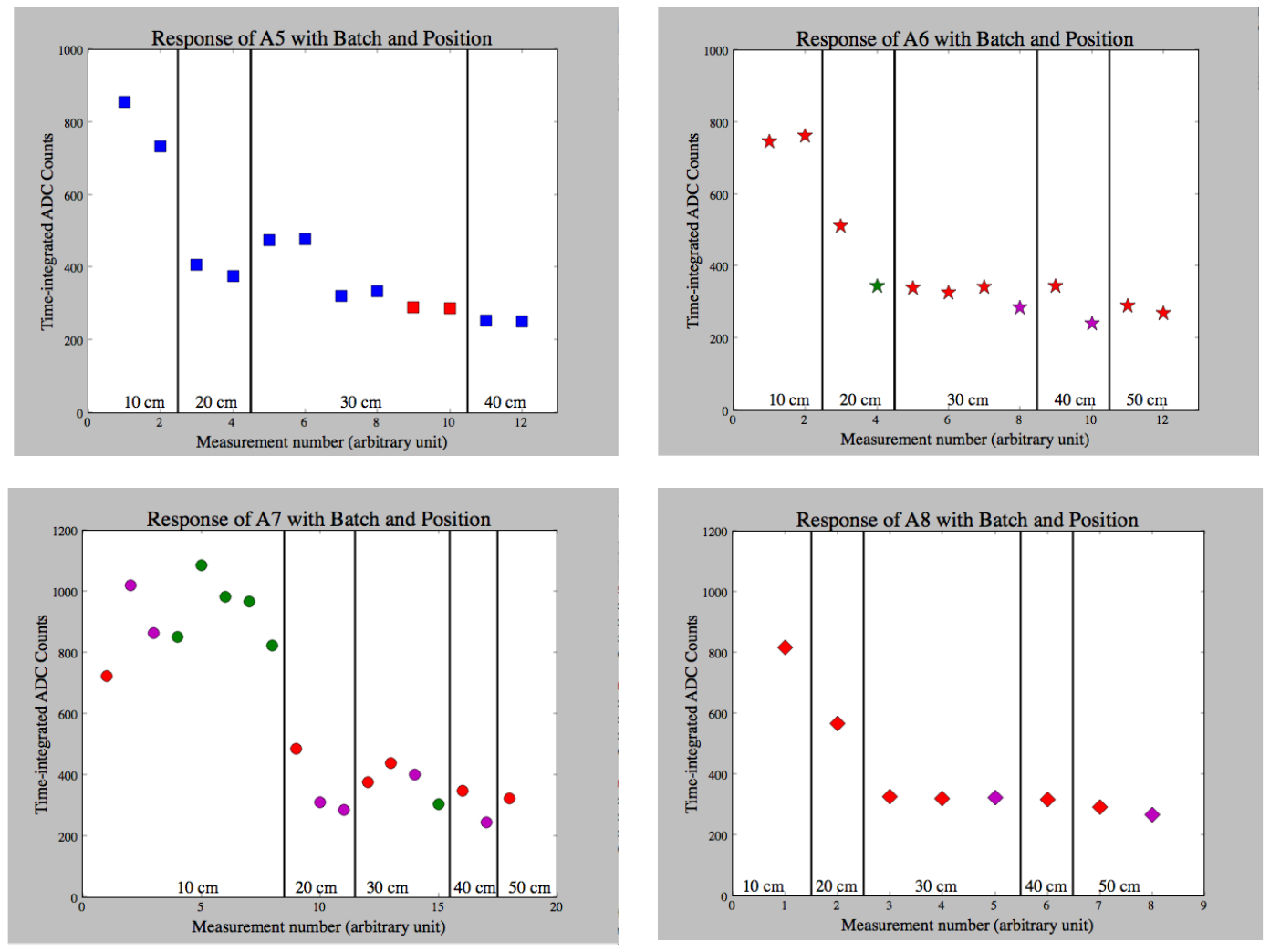}
\caption{(color online). 
Plots corresponding to measurements of integrated
charge, $Q_{tot}$, of early light for four lightguides.   Measurements were
made across four batches of LAr indicated by the color of the symbol.
Measurements are made at multiple locations along a bar.\label{fourbars}
}
\end{figure}

The studies presented here involve four batches
of LAr and four bars, and were performed over about 60~days in order
to study a variety of sources of variation of response.

Fig.~\ref{fourbars} presents measurements for each of the four light guides
under study, with a different symbol associated with each bar.  
The colors indicate the LAr batch in the test stand for a given 
measurement, where the time-order was:  blue, red, magenta and green.   
Measurements are taken at up to five locations along the bar (10, 20,
30, 40 and 50 cm), and often multiple times.    The source is removed
and replaced between every measurement, even in the case where a
measurement is taken multiple times at the same location.  The exact
location is varied within $\sim 1$ cm.  The purpose of this is to
sample multiple areas of coating at each distance.    Within a given 
location indicated on the plots, the measurements are presented in time
order.   Every measurement has more than 70,000 entries, and so the 
error bars are not visible on the plots.

\begin{figure}[tb]
\centering
\includegraphics[width=0.85\textwidth]{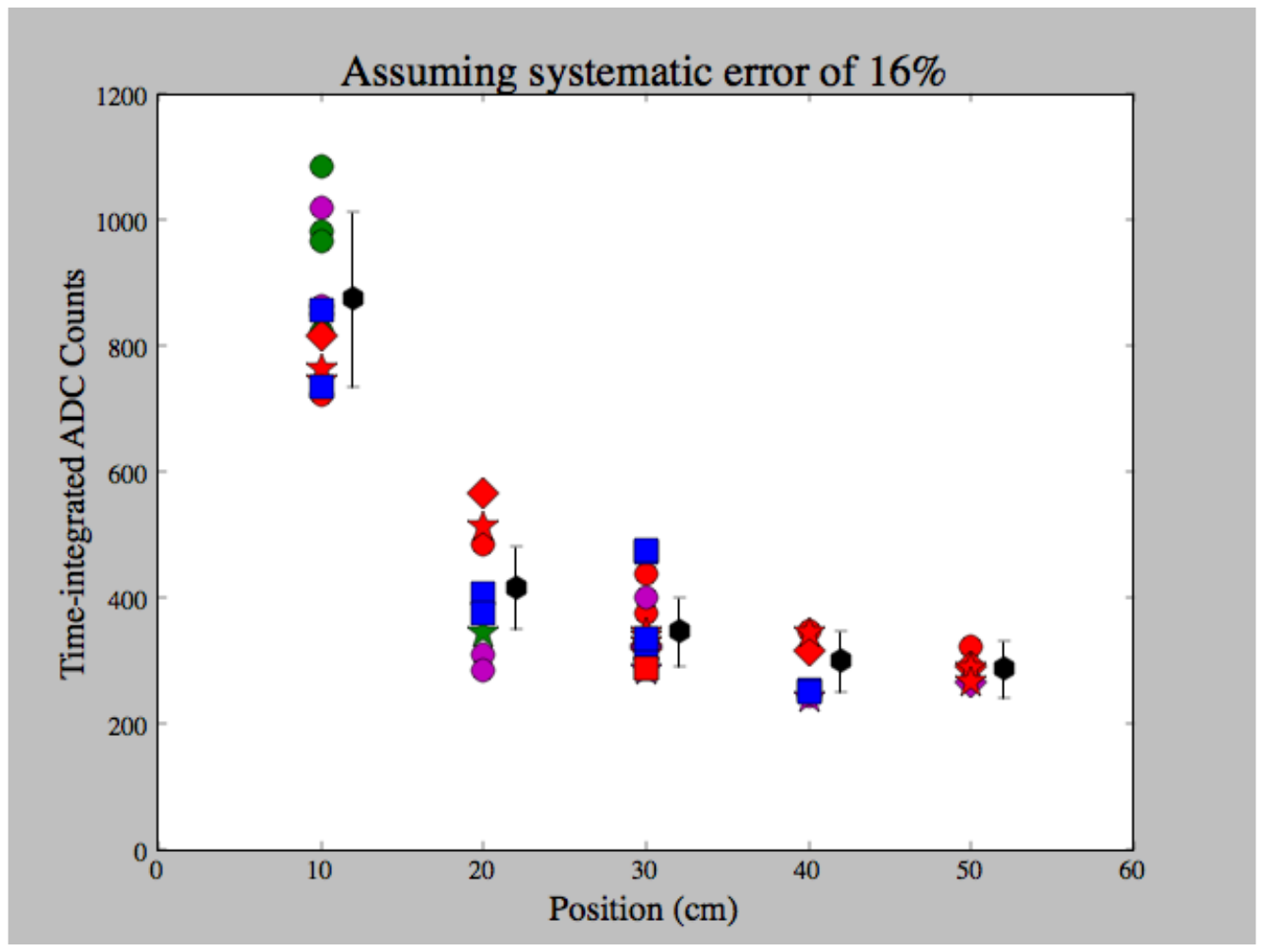}
\caption{(color online). 
$Q_{tot}$ as a   function of position for four batches
  (colors) and four bars (symbols).  Black
hexagons indicate the average and error bars indicate a 16\%
systematic spread. \label{fourbarsoneplot}
}
\end{figure}

Fig.~\ref{fourbarsoneplot} transfers information from
Fig.~\ref{fourbars}, maintaining the meanings of the symbols for each
bar and colors for each batch.  In this figure, time information has
been removed and the time integrated ADC counts, $Q_{tot}$, are
plotted as a function of location of the source.  The mean at each
location is indicated by the black hexagon with the error bar. 
The deviation of measurements from the mean is 
such that a 16\% systematic spread encompasses 68\% of the data points. 

The following conclusions can be drawn from Fig.~\ref{fourbars} and
Fig.~\ref{fourbarsoneplot}:
\begin{itemize}
\item At a given source location, the spread of measured values
  is very large compared to the statistical error.
\item The spread in measurements at each location is, to a good approximation, the
  same fraction (16\%) of the measured value.
\item The spread is not due to LAr batch variations.  The
  results from multiple batches are quite consistent within the
  spread.
\item There is no evidence of systematic degradation with time due to
  an external source such as UV light.
\end{itemize}
The spread in measurements seems likely to be due to variations
in the quality of the coating within the region sampled at each location.

\begin{figure}[tb]
\centering
\includegraphics[width=0.85\textwidth]{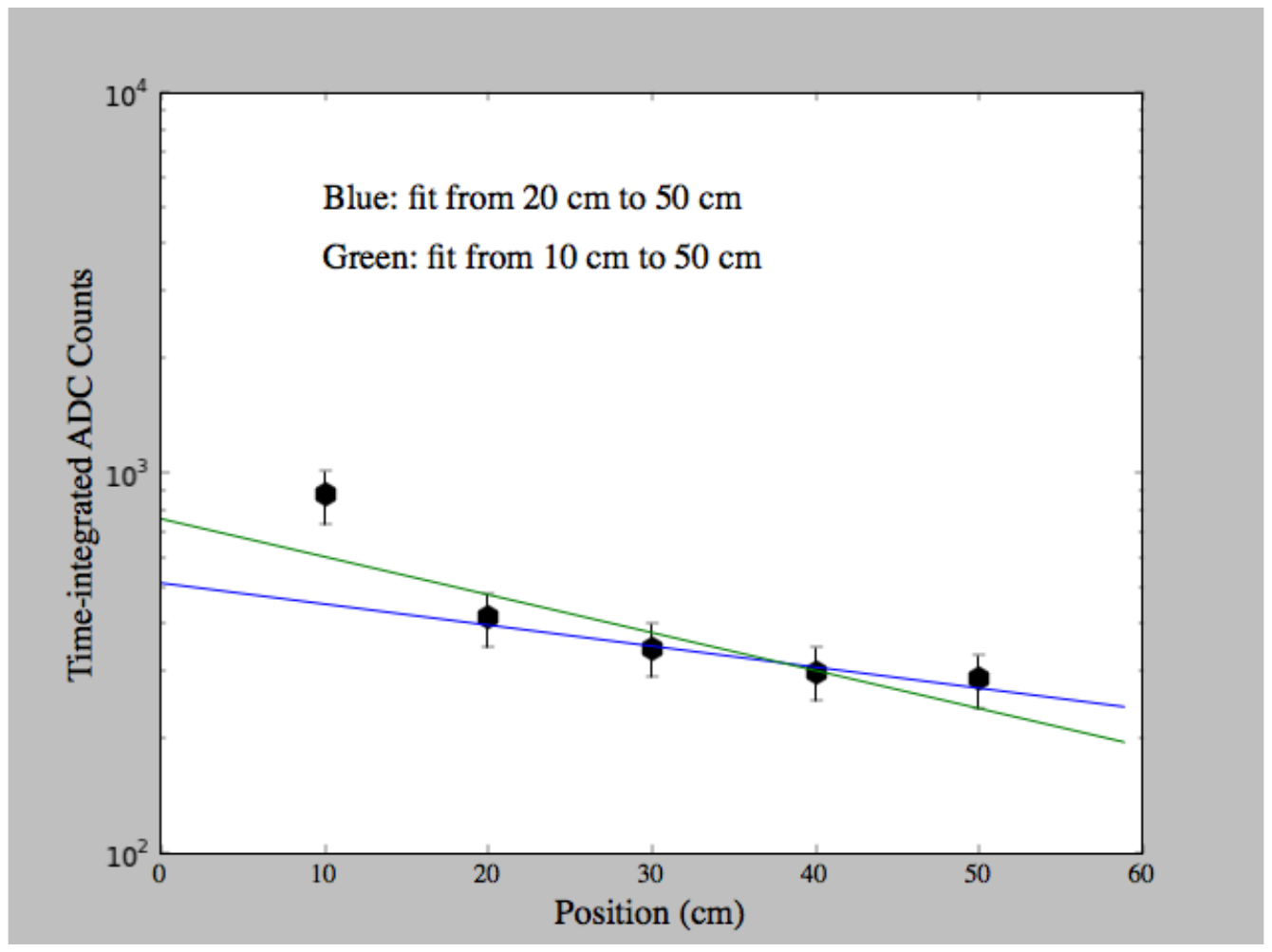}
\caption{(color online). 
Exponential fits to the mean measurement at each position,
  taking the error to be 16\%.   The results at 10 cm are far from the
expectation for an exponential, whether this point is included 
(green) or excluded (blue) in the fit.  
\label{means}
}
\end{figure}

In principle, one expects an exponential attenuation along the bar.
Fig.~\ref{means} presents the mean of the measurements in LAr (black
points) as a function
of location along the bar on a semi-log plot.   The point at 10 cm lies 
significantly higher than the expectation for an exponential.  
The
green curve, which includes the 10 cm point in the fit, results in an 
attenuation length of 44 cm, while the blue curve, which excludes the 
10 cm point,  indicates an
attenuation length of 79 cm.   

These results are in qualitative agreement with the warm, uncoated 
acrylic bar measurements presented in 
Sec.~\ref{attennocoat}),  where the overall fit gave 38$\pm$1~cm
attenuation length.   It is possible that the observed deviations can be understood by considering a ray tracing model, described in Sec. \ref{sec:attensim}, which indicates non-exponential attenuation and is a better match to this data.

To establish the performance of the coating, apart from the 
attenuation, we use the 10 cm point as a benchmark.       This is likely to be a
conservative
estimate as some attenuation may be occurring over the 
10 cm.
The $Q_{tot}$-to-PE conversion  yields 21.9 PE (7.2 PE)
at 10 cm (50 cm).  This result can be compared with the previously
reported light guides that reported $\sim 7$ PE at 10 cm. 

\section{Interpretation of Attenuation Data} \label{sec:attensim}

\begin{figure}[tb]
\centering
\includegraphics[width=0.6\textwidth]{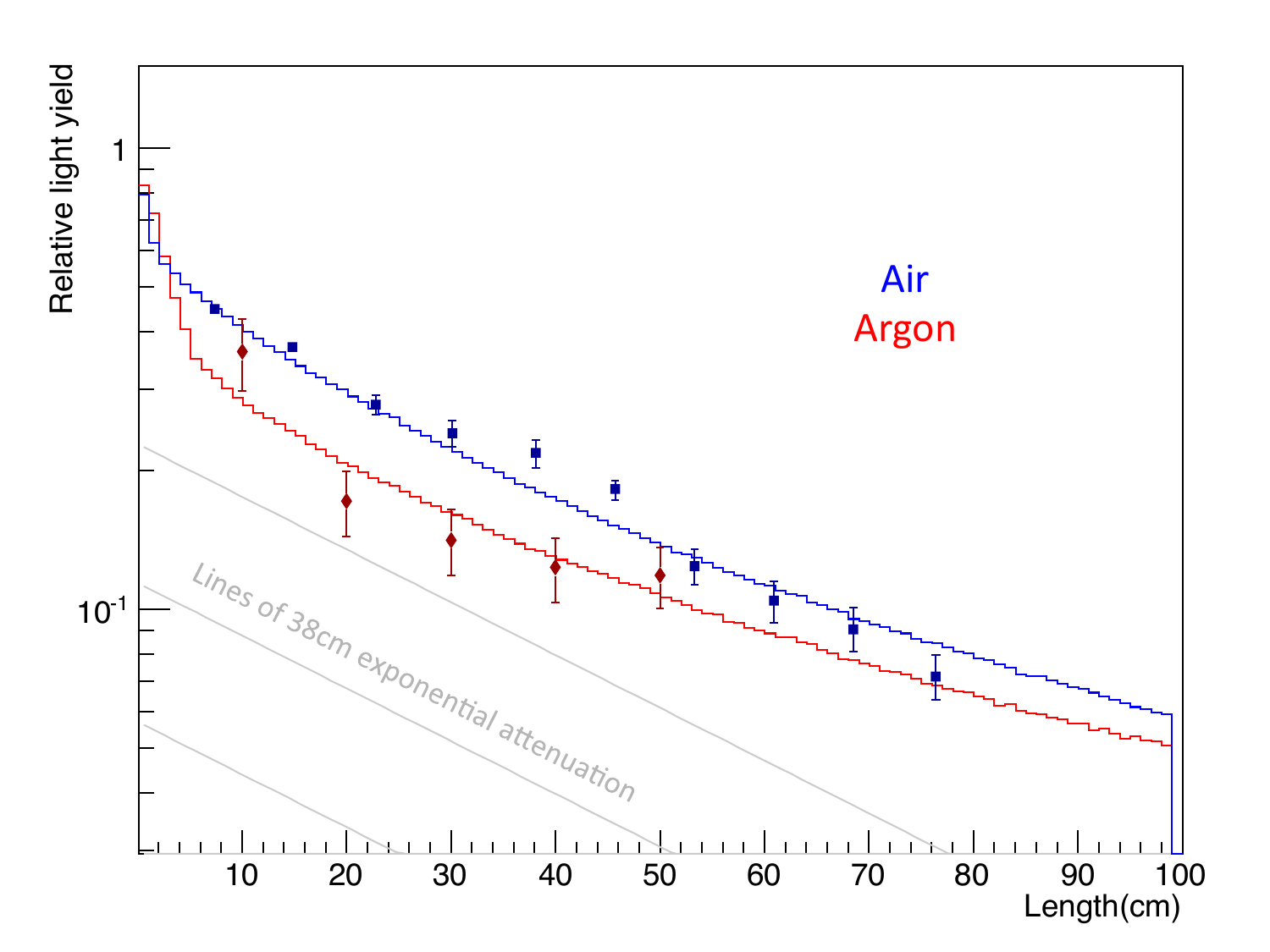}
\caption{(color online) Calclated attenuation curves using the model described in the text.   The coefficient of surface absorption is determined to be 3\% per reflection within this model and is used to make a prediction on the argon data.
\label{fig:NonExpAtten}
}
\end{figure}

In order to interpret the attenuation data from lightguides operated air and liquid argon described in this paper, a simple ray tracing model was developed \cite{ben_atten_sim}. In this model, isotropic light rays emanating from the TPB surface are produced and attenuated according to partial reflection at the dielectric boundary and absorption at imperfect optical surfaces. The model has a single free parameter, the coefficient of surface absorption, which is tuned on data taken in air. The tuned model is then compared to data for liquid argon. This comparison can be seen in figure \ref{fig:NonExpAtten}. The model is in qualitatively better agreement to the liquid argon attenuation data than a simple exponential fit. Non-exponential attenuation is present due to the changing angular distribution of rays along the bar, with the higher angle rays being attenuated more quickly than those which are more parallel to the light guiding direction. Different behavior is expected for light guides operated in air and liquid argon, due to the different refractive indices of the two environments.

The effects of surface roughness were also incorporated into the model. The effects were found to be insignificant in all cases where surface undulations occur on scales longer than the wavelength of the guided light. This is because for all reasonable values of angular non-uniformity, only a small fraction of rays are diverted from the supercritical to subcritical reflection regions. Roughness on scales shorter than the wavelength of light would appear as a contribution to the surface absorption coefficient, which is tuned to data in this model.

The effects of wavelength dependent attenuation were investigated. Monochromatic attenuation lengths measured in the same apparatus described in section [Stuarts section] with different colored LEDs were used to tune the surface absorption coefficient as a function of wavelength. Averaging over the TPB emission spectrum, the predicted attenuation curves showed a deviation from the assumed monochromatic attenuation which was insignificant when compared with experimental errors. More information about this model and its predictions can be found in \cite{ben_atten_sim}.

\subsection{Comparison of Acrylic vs Polystyrene Coatings.}

\begin{figure}[tb]
\centering
\includegraphics[width=0.85\textwidth]{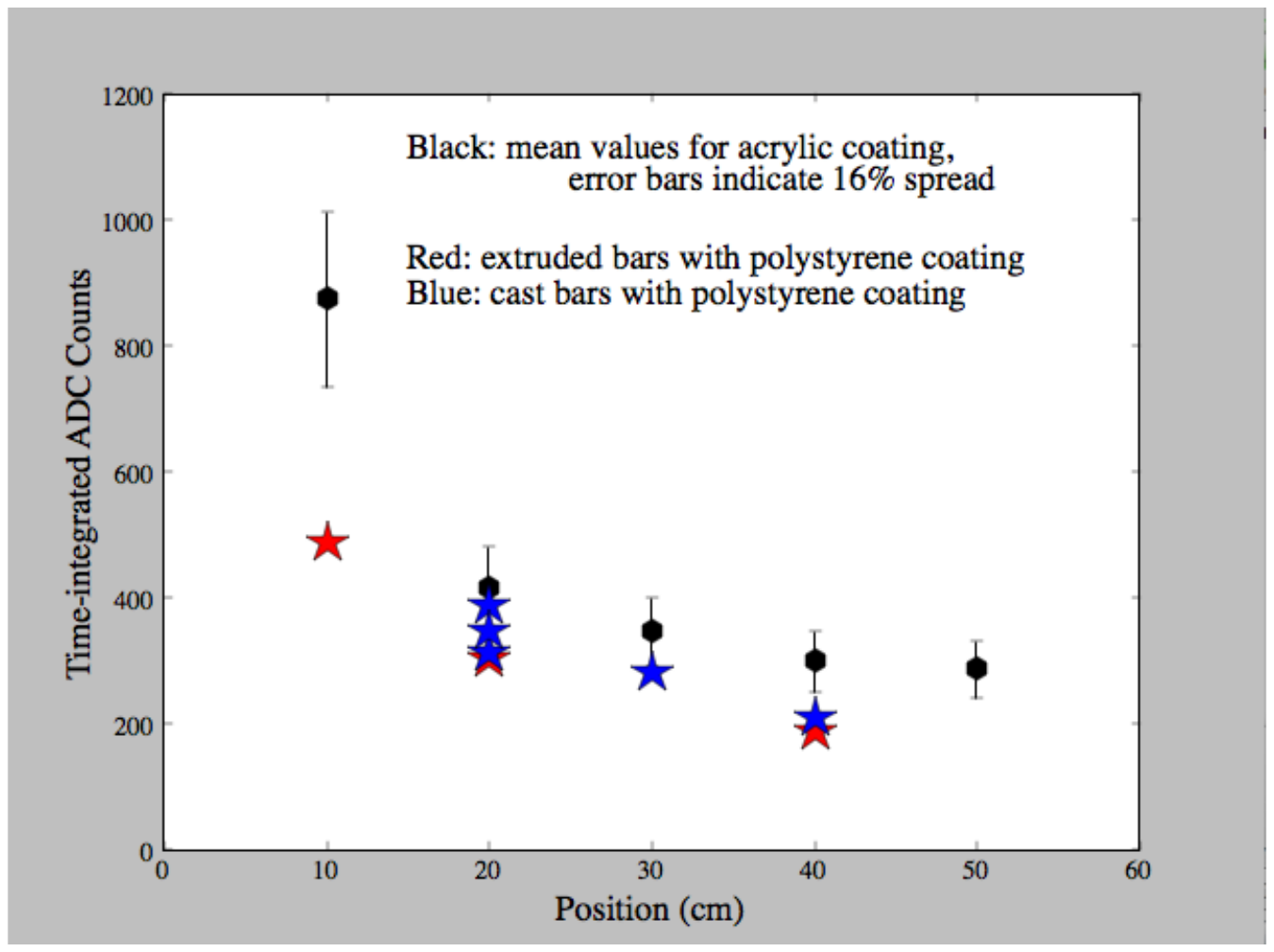}
\caption{(color online). 
Comparison of LAr measurements for two coatings.    
Black circle:  mean and spread of data from acrylic coating on cast acrylic bars.   
Red star: measurements on extruded bars with polystyrene coating.
Blue star: measurements on cast bars with polystyrene coating. \label{psbars}
}
\end{figure}

The cast bars with UVT33\% coating are compared to cast bars and 
extruded bars with PS25\% coating in Fig.~\ref{psbars}.
The acrylic coating is consistently better than the polystyrene coating by
a factor of about 1.3.    

\section{Conclusions\label{sec:conclusion}}

This paper has presented information useful to simulating lightguide
paddles in future LArTPC experiments.  
We have presented a comparison of response of lightguides prepared
with two coatings, UVT33\% and PS25\%.  We find that the response
levels are similar for lightguides in LAr, although the transmission
tests at room temperature indicated that PS25\% was the better coating.
At 10 cm, typical response of the light guides is about 21 PE, which
is a factor of three improvement over previously reported lightguides. 

\acknowledgments

The authors thank the National Science Foundation (NSF-PHY-084784) and 
Department Of Energy (DE-FG02-91ER40661). 
We thank Dr. A.  Pla of Fermi National Accelerator Laboratory for helpful
discussions and use of equipment.

\bibliographystyle{plain}

\bibliography{coatings_jinst}

\end{document}